# de Broglie waves as the "Bridge of Becoming" between quantum theory and relativity

Ruth E. Kastner

Foundations of Physics Group

University of Maryland, College Park, USA

7/5/11

ABSTRACT: It is hypothesized that de Broglie's 'matter waves' provide a dynamical basis for Minkowski spacetime in an antisubstantivalist or relational account. The relativity of simultaneity is seen as an effect of the de Broglie oscillation together with a basic relativity postulate, while the dispersion relation from finite rest mass gives rise to the differentiation of spatial and temporal axes. Thus spacetime is seen as not fundamental, but rather as emergent from the quantum level. A result by Solov'ev which demonstrates that time is not an applicable concept at the quantum level is adduced in support of this claim. Finally, it is noted that de Broglie waves can be seen as the "bridge of becoming" discussed by Elitzur and Dolev (2005).

1. Introduction.

This paper offers a proposal based on a relational or antisubstantivalist view of spacetime.[1] Its main aim is not primarily to defend the relational view, although I will briefly review that view and some basic arguments for its plausibility, to set the stage for the specific claim presented: the de Broglie "matter wave" is ontologically prior to spacetime, and the relationship between the former's phase and group velocities can be seen as the ontological basis of the structure of Minkowski spacetime.

---

[1] More precisely, the view presented here is relationalism about ontology: i.e., that spacetime does not exist as a substance. See, e.g., Skow (2007). For a comprehensive account of the relational view, see Sklar (1974).

The relational or antisubstantivalist interpretation of spacetime is concisely described by Weingard (1977):

"According to the relational theory…what we call "space" is simply the totality of actual (and perhaps possible) spatial relations between material objects and/or concrete material events. If there were no material objects and concrete material events, space would not exist, for the relata of the relations constitutive of space would not exist, just as a family tree cannot exist without there being people to bear the family relations to each other that are constitutive of the tree." (p. 167)

In this view, the motion of an object is taken to be based on its relations with other observable objects or phenomena, and not on an absolute trajectory through an independently existing spacetime substance.

2. de Broglie's proposal and its significance

The "matter waves" hypothesized by Louis de Broglie (1923) have been well-confirmed in numerous experiments. By the term 'matter wave' we intend exactly what de Broglie intended: i.e., the assumption that ontologically real wavelengths are associated, not only with massless particles (photons) but also with particles of finite mass including atoms, molecules, and even Buckeyballs (see, e.g. interference experiments by Arndt and Zeilinger (2003). For example, the Davisson and Germer experiment (1927) obtained a diffraction pattern when sending electrons through a double-slit apparatus. Such patterns only appear if an entity possessing wavelength is physically associated with the measured system(s). One can deduce the wavelength of the entity by examining the interference pattern produced in any given experiment.[2]

It must be emphasized that, despite the connection of de Broglie's postulated 'matter wave' with the historical development of nonrelativistic quantum mechanics, it is fundamentally relativistic in nature. Its relativistic nature arises from the fact that it ascribes an intrinsic frequency to a material particle based on the latter's relativistic mass, as equivalent to its energy.[3] This provides a deep link between quantum mechanics and relativity which merits further exploration, and offers an opportunity to make progress in reconciling the two theories.

---

[2] It should be noted that the Bohmian theory (cf. Holland 2000) separates the 'guide wave' from the postulated pointlike Bohmian particle. In that case, the 'matter wave' refers to the guide wave,. But any possible conflict or inconsistency between the current proposal and the Bohmian approach—which is just one proposed interpretation--- should not be considered reason to reject the current proposal.

It must also be noted that the wavelike aspect appears in solutions to quantum mechanical wave equations whose home (in the case of the Schrödinger equation) is configuration space, not spacetime. That is, consider a system of N particles. Their associated de Broglie wave is defined with respect to a 3N-dimensional configuration space; an independent degree of freedom is assigned to the $x,y,$ and $z$ coordinates for *each* particle. So the 'matter waves' cannot be thought of as literally 'propagating in spacetime' which has only three ($x,y,$ and $z$) spatial degrees of freedom. However, because specific wavelike phenomena attributable to these solutions *do* appear in spacetime (i.e., diffraction patterns), clearly there is some ontological relationship, or at least connection, between the 'matter waves' and spacetime. It is this relationship which is examined herein.

Despite universal acceptance of the de Broglie wavelength $\lambda = h/|p|$ (where $|p| = \gamma m_0 v$) is the absolute value of its 3-momentum) as a real physical quantity, or at least as a quantity able to be indirectly observed as discussed above, de Broglie waves possess a superluminal phase velocity $u = c^2/v$, usually regarded as "fictitious". According to de Broglie's theory, for a particle at rest ($v=0$), the phase wave aspect propagates at infinite velocity, which seems unphysical. Instead, it is customary to associate the "real" or physically acceptable part of the wave with the group velocity $v$.

While it is true that the aspect of the wave propagating with the superluminal or even infinite phase velocity carries no energy and thus cannot be detected, it is suggested here that its existence should not be dismissed. For one thing, in order for any wave to have a physically real group velocity component capable of giving rise to detectable phenomena, it must have an underlying phase velocity. That is, you can't just accept the group velocity and reject the phase velocity: they come as a complete ontological package—even if the ontology is that of an entity that cannot be thought of as being fully contained in spacetime.[4]

---

[3] De Broglie waves are the basis for relativistic wave equations (such as the Klein-Gordon equation) as well, although they cannot, according to the usual analysis, be given a single-particle interpretation. (One way to do so is by interpreting the K-G equation as describing a particle moving both forward and backward in time.)

But the more important reason to accept the ontological reality of the superluminal phase waves is that they provide a dynamical basis for the apparent structure of Minkowski spacetime, as will be explicated in what follows. In pursuing this possibility, it is worthwhile to recall Mach's view that space and time would not exist without matter. Elitzur and Dolev (2005) argue for a Machian-based spacetime picture of 'Becoming,' in which, in their terms, "The spacetime of every reference frame is formed by the gravitational/electromagnetic interaction of that frame with its environment. These interactions, which occur in the pre-spacetime stage, determine the spatio-temporal distance between events..." (p. 206). The title of the Elitzur-Dolev paper is "Becoming as a Bridge Between Quantum Mechanics and Relativity." While the process of the emergence of spacetime is viewed a little differently here,[5] I will argue below that de Broglie waves can be seen as the mechanism behind Elitzur and Dolev's 'Bridge of Becoming.'

3. Main argument

It should first be emphasized that de Broglie himself first noted the coincidence of his oscillation's phase and group aspects with the Minkowski spatial and temporal axes respectively (de Broglie's dissertation (1925), section 1.3, "Phase Waves in Spacetime" ). However, the prevailing substantivalist view of spacetime, as expressed by numerous authors[6], describes de Broglie waves for a particle in motion as an *effect* of the relativity of simultaneity.[7] Recall that de Broglie's postulate is that a particle of rest mass $m_0$ is associated with an oscillation of frequency $f = m_0 c^2 / h$, which takes the form of a standing wave of constant phase at all spacetime points for a particle at rest. In the standard, substantivalist account, this oscillation is seen as passively 'riding along' with those points. The oscillation's transformed properties when viewed from a moving frame (with a velocity $v$ with respect to the particle's rest frame) are then taken to be dictated by the *a priori* structure of spacetime. Those properties include the appearance of a finite de Broglie wavelength for the particle as seen from the moving frame, based on the fact that a particular phase is associated with a set of points which are simultaneous in the rest frame but not in the moving frame (See Baylis 2005 for a particularly clear account).

---

[4] A case in point is the Aharonov-Bohm effect, in which the electromagnetic vector potential, previously considered not physically real, was shown capable of giving rise to observable effects (Aharonov and Bohm, 1959).
[5] The specifics of this process will be explored in a separate paper.
[6] E.g.. Baylis (2005), Rindler (2006 , p, 121).
[7] It should be noted that the subtantivalist view is subject to sustained criticism by Harvey Brown (cf. Brown 2005).

However, it does not seem to have been noticed that one can readily invert the cause/effect arrow of the usual description, and argue instead that *the relativity of simultaneity is an effect of de Broglie waves*, together with a relativity postulate. This will be demonstrated in what follows.

Consider the following postulates:

Postulate 1: all material entities have an oscillation associated with them, of frequency $f = \gamma m_0 c^2 / h$. For a particle with $p = 0$, this oscillation takes the form of a standing wave of constant phase at all possible spacetime locations[8] (de Broglie's postulate).

Postulate 2: The value of $p$ is relative to the observer (the principle of relativity, stated dynamically).

This is now all you need to obtain the appearance of a Minkowski spacetime "theater of experience" (Figure 1). With respect to the lab frame S,[9] a particle P propagates at velocity $v$ which is the group velocity of its associated de Broglie wave (Postulate 1). Note that this serves as the particle's time axis (labeled $t'$). From our perspective in the lab, the particle's hyperplane of constant phase appears not to be constant, by Postulate 2 (since P is moving relative to us and our own oscillations are taken as defining the hyperplane of constant phase or simultaneity). Instead, it has a finite and physically detectable wavelength of $\lambda = h/|p|$. This wavelength expresses the relativity of simultaneity and, it is suggested here,

---

[8] The term "possible spacetime locations" is intended to reflect that such 'locations' are defined only relationally; a location does not exist as a 'spacetime point.' So, for example, two observers at rest with respect to the particle and separated by any distance would measure the same phase.

[9] In using the term "lab frame" here, it is not implied that there is some independently existing substance corresponding to this "frame." It is merely a perspectival frame of reference.

*is its true dynamical origin*, rather than being the effect of a pre-existing simultaneity relation considered, in the usual way, as an essential property of spacetime.

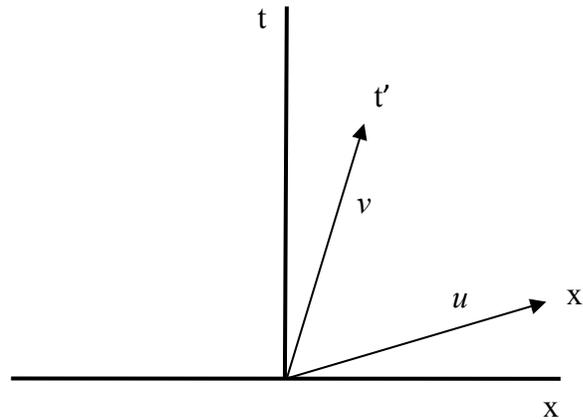

Figure 1. de Broglie phase waves, travelling at velocity *u* relative to an observer, serve to define the spatial hyperplane *x′* for a quantum of matter moving at apparent group wave velocity *v*. The group wave component defines the quantum's temporal axis *t′*.

Note also that the wave's phase component with velocity *u,* as measured by us (indirectly through its detectable velocity *v*), serves as P's spatial axis relative to ours: that is, as its plane of constant phase and therefore constant *t ′* , by Postulate 1. It is suggested here that this is more than a convenient coincidence: i.e., that the *x′* axis is not just 'sitting there' waiting for a de Broglie wave to coincide with it. Rather, we should reinterpret what is usually presented as the moving particle's primed spatial axis, given *a priori* by "the structure of Minkowski spacetime," simply as its de Broglie phase wave traveling at velocity *u > c*. *The spatial axis is no more, and no less, than the de Broglie phase wave; and the temporal axis is no more, and no less, than the de Broglie group wave*. We now have the dynamical "horse" before the phenomenal spacetime "cart," instead of the other way around.

Some additional comments are probably in order regarding the conceptual coherence of this account, which no doubt seems quite radical given the standard assumption of a pre-existing Minkowski spacetime and the traditional prohibition on superluminal propagation. It should be kept in mind that the usual depiction of a particle in terms of a Minkowski spacetime diagram, and its analysis via our calculational tools based on the spacetime construct, necessarily "smuggle in" the notion of a substantial spacetime, whether it actually exists or not; and the constraints of consistency demand that its de Broglie phase oscillation appear to 'propagate' at infinite velocity. But it is claimed here that this depiction of the wave as propagating "in spacetime" is an artifact of our imposition of the spacetime construct on the underlying dynamical object(s), and that the latter should be considered the ontological supports of that construct, rather than the other way around. After all, as noted earlier, the oscillation takes place in a 'higher space' described by configuration space.[10]

Furthermore, recall that the uncertainty principle tells us that in the case of a particle with definite momentum, we can have no knowledge of a particle's position; or, put in ontological terms, that the particle does not *have* a definite position. This is consistent with the idea that, from the particle's point of view, it is 'everywhere at once.' Taking this ontology seriously implies that spacetime, and the attendant notion of 'location of a particle' is not fundamental. Although de Broglie himself did not make the explicit move to antisubstantivalism, he clearly noted the futility of trying identify a particle with a spacetime location:

> Must we suppose that this periodic phenomenon occurs in the interior of energy packets? This is not at all necessary; the results of §1.3 will show that it is spread out over an extended space. Moreover, what must we understand by the interior of a parcel of energy? An electron is for us the archetype of isolated parcel of energy, which we believe, perhaps incorrectly, to know well; ….That which makes an electron an atom of energy is not its small volume that it occupies in space, I repeat: it occupies all space, but the fact that it is undividable, that it constitutes a unit. (de Broglie 1925, 8)

Another way of understanding the claim presented herein is that Minkowski spacetime is a very useful map which captures phenomenal relationships between dynamical entities (in particular, relativistic invariance). But it only that: a map, which is not equivalent to the actual substantial reality supporting those phenomena. We may be able to locate a city on a map for empirical purposes, but that does not mean that the city is literally contained in the map. So, rather than dismiss the notion of an "infinite (or superluminal) velocity" as unphysical, we should keep in mind that *the appearance of such an "unphysical" propagation is likely an artifact of our imposition of a spacetime picture on an entity which transcends that depiction.* The apparently 'unphysicality' of the superluminal propagation can be considered conceptually analogous to what might be considered an 'unphysical' trajectory from one discontinuous portion to another on the Goode Homolodine Projection ("orange peel map"), in which areas are preserved by making the map discontinuous. The latter captures important metric features of the space it portrays, but if mistakenly taken as equivalent to the portrayed reality, needlessly suggests 'unphysicality.'

---

[10] Indeed, for two or more quantum entities, the associated wave 'lives' in a multidimensional configuration space; such a wave does not literally 'propagate in spacetime'. We describe here the limit in which a propagating system can be described by a single-particle momentum eigenstate, which is the limit in which classical relativity (which presupposes a definite velocity in spacetime) applies. While it is a matter for future study how the account can be explicitly extended to multi-quanta states, we see no reason *a priori* to rule out the present account as potentially applicable to such states, with the assumption that spacetime is an epiphenomenon of a more fundamental entity described by Hilbert space.

3. Further Implications.

Note that the separation between the group and phase velocities is a direct result of the dispersion relation embodied in the dependence of frequency on momentum for particles with nonzero rest mass $m_0$:

$$\hbar^2 \omega^2 = \hbar^2 k^2 c^2 + m_0^2 c^4. \tag{1}$$

While the phase velocity $u$ is given simply by

$$\omega/k = \gamma m_0 c^2 / \gamma m_0 v = c^2/v, \tag{2}$$

the group velocity, $v$, is given by the dispersion relation formula:

$$v = d\omega/dk \tag{3}$$

and by differentiation of (1),

$$2\omega \, d\omega = 2 c^2 k \, dk,$$

$$d\omega/dk = c^2 k/\omega = v. \tag{4}$$

For massless particles like photons, there is no such separation of the phase and group velocities, since setting $m_0$ to zero in (1) gives simply $\omega = kc$, and thus $u = v = c$.

This interpretation of the de Broglie phase and group velocities as the dynamical foundation of the structure of Minkowski space provides the following interesting perspective: it is matter (i.e., nonzero rest mass) and its attendant de Broglie wave that *creates* the apparent separation between the temporal and spatial axes of spacetime (see Figure 2). If there were no finite rest mass, there would be no separation; everything would propagate on null cones; no time would pass and no space would be traversed. The rest mass dispersion effect can be seen as a direct cause of the differentiation of the temporal and spatial dimensions of experience out of a dimension (the null cone) in which they are undifferentiated.

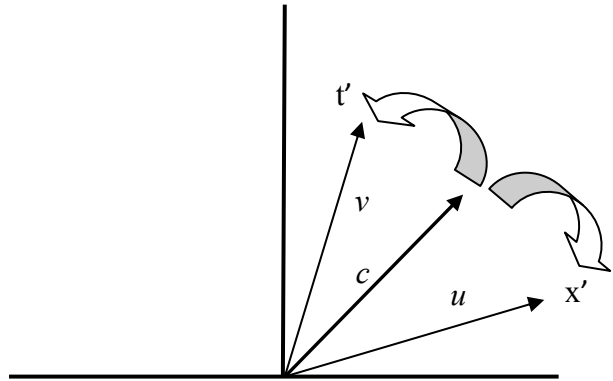

Figure 2. The dispersion relation for finite rest mass creates the separation of spatial and temporal axes in Minkowski spacetime.

t

In addition, it is interesting to note that the dispersion relation (4), together with the quantum mechanical relationship of energy to frequency and momentum to wave number of material particles, leads to one of Hamilton's equations:

$$dx/dt = d\omega/dk = \partial H / \partial p \qquad (4a)$$

and, in the light of the previous discussion, one can 'read off' the differentiation of spatial and temporal axes on the left-hand side of (4a) in terms of the dispersion relation. In natural units ($c=h=1$), the left-hand side is unity for massless particles such as photons. In the limit in which the two parameters $x$ and $t$ describe the same physical entity, the quantity $dx/dt$ must be unity. If that entity is allowed to 'split' into two distinct physical entities, $dx/dt$ will be allowed to deviate from unity; thus any deviation from unity corresponds to the appearance of finite mass for which $dx/dt \neq 1$.

Further support for the idea that space and time are emergent, as opposed to fundamental, is found in recent work of Solov'ev (2010). He has shown in the context of atomic collision theory that the concept of time is not unambiguously definable for impact energy comparable to the splitting of energies in the different inelastic channels (quantum states) of index $n$. Time can be unambiguously defined only if one neglects the dependence of the internuclear radial momentum $P_n(R)$ on $n$, which is not possible for small $n$ and low impact energy $\epsilon$. He observes that "at low impact energy, the solution of the [time-independent] Schrodinger equation is the superposition of nuclear motions in different n-channels having significantly different momenta $P_n(R)$, and unique time does not exist." (Solov'ev 2010, p. 6.)

Specifically, only if the energy of the system $\epsilon$ is much greater than the energy associated with each n-channel $E_n$ can one make the approximation (Solov'ev 1995, p.471):

$$P_n(R) = MV(R) + E_n(R)/V(R), \qquad (5)$$

Where $M$ is the nuclear mass and $V(R)$ is the nuclear radial velocity in this limit (obtained from the leading term in the expansion of $P_n(R)$ with respect to small $E_n(R)$ and which, in the classical theory has the meaning of the Coulomb radial velocity [12]).

---

[12] $V(R)$ is defined in this limit as $(2M\epsilon)^{1/2}/M$. See Solov'ev (2009), eq. 16.

In the limit in which the second term can be neglected, and multiplying by $dR$, one can then say

$$M \, dR / P_n(R) \approx dR/V(R) \approx dt, \qquad (6)$$

but this time differential can only be said to exist in the above limits, corresponding to the classical domain. In view of this result, Solov'ev argues that quantum entities are properly understood as transcending the spacetime construct, a view which is harmonious with the interpretation presented here.

Finally, another argument against the idea of spacetime as a pre-existing substance (i.e. prior to or more basic than rest mass) is the following. It is commonly supposed that it is the "geometry of spacetime" that confines photons to geodesics and which dictates the precise speed of light. Recall the "mystical formula" of Minkowski which defines the relationship of the temporal dimension to the spatial dimension(s)[13]:

$$3 \times 10^5 \text{ km} = i \text{ sec} \qquad (5)$$

Yet this alleged "intrinsic geometry" of spacetime is not exact: there is a finite possibility for a photon propagation at speeds greater or less than $c$. Physically, this arises because of virtual particle production in the vacuum. (See, for example, the Scharnhorst effect.[14]) This indicates that quantum mechanics is more fundamental than theories of spacetime, and that the relation (5) is only approximate. It is ultimately dependent on the behavior of photons and virtual particles, which are quantum mechanical objects. It might be objected that a quantum theory of gravity would take all this into account. But the point remains that such a theory would admit an alteration of the supposed 'intrinsic' properties of spacetime based on quantum entities. This strengthens the case for the phenomenal dependence of spacetime on more fundamental dynamical entities such as the de Broglie wave.[15]

---

[1] H. Minkowski (1908).
[14] G. Barton, K. Scharnhorst (1993)
[15] It should perhaps be noted here that the relevance of gravity is obviously deeper. However, general relativity lies outside the framework of the present paper. It is suggested that the connection between de Broglie waves and general relativity constitutes unexplored and potentially very fruitful territory.

4. Conclusion.

It has been argued that the Minkowski spacetime "theater of events" is an epiphenomenon of material quanta arising from their associated de Broglie waves, along with the principle of relativity interpreted dynamically. The "spacetime" in which a quantum of matter appears to propagate arises as a result of de Broglie phase and group waves, the temporal axis being identified with the group wave component and the spatial axis being identified with the phase wave component. Thus "spacetime" should not be thought of as a substance, but only a way of describing observable effects arising from the existence of finite mass. A result by Solov'ev showing that time is not well-defined in the limit of small impact energies for atomic collisions supports this interpretation, insofar as it confirms that 'time' is not an applicable physical parameter for situations far from the classical limit.

Therefore, in an account in which de Broglie waves are the dynamical basis for the differentiation of spatial and temporal dimensions by way of the dispersion relation for nonzero rest mass, they can be seen as the "bridge of becoming" discussed by Elitzur and Dolev (2005). That is, they give rise to the phenomenon of relativistic spacetime out of a quantum-mechanical "pre-spacetime" realm, and thereby provide a crucial connection between quantum theory and relativity.

Acknowledgements

The author is grateful to Avshalom Elitzur , Eugene Solov'ev, and anonymous referee for valuable comments.